\documentstyle[preprint,aps]{revtex}

\begin{document}
\draft
\preprint{KANAZAWA 94-15,  
\ July 1994}  
\title{
Polyakov loops and monopoles in QCD
}
\author{
Tsuneo Suzuki$^{\small a}$
\footnote{ E-mail address:suzuki@hep.s.kanazawa-u.ac.jp},
Sawut Ilyar$^{\small a}$, 
Yoshimi Matsubara$^{\small b}$
\footnote{ E-mail address:matubara@hep.s.kanazawa-u.ac.jp},
Tsuyoshi Okude$^{\small a}$ \\ 
 and Kenji Yotsuji$^{\small a}$
}
\address{$^{\small a}$
Department of Physics, Kanazawa University, Kanazawa 920-11, Japan
}
\address{$^{\small b}$
Nanao Junior College, Nanao, Ishikawa 926, Japan
}
\maketitle

\begin{abstract}
Monte-Carlo simulations of abelian projection of 
$T \neq 0$ pure lattice QCD show that
1)\ Polyakov loops written 
in terms of abelian link fields alone play a role of an order parameter of 
deconfinement transition, 
2)\ the abelian Polyakov 
loops are decomposed into contributions from Dirac strings of monopoles 
and from photons, 
3)\ vanishing of the abelian Polyakov loops 
in the confinement 
phase is due to the Dirac strings alone and 
the photons give a finite contribution 
in both phases. 
Moreover, these results appear to hold good with any 
abelian projection as seen from the studies in the maximally abelian gauge and 
in various unitary gauges. 
\end{abstract}

       
\newpage

\section{introduction}
Color confinement mechanism in QCD is still to be understood.
Many works\cite{suzu93,shiba1,shiba2,shiba3,cosmai,yee,polika,wensley,shiba4,shiba6,ejiri,stack,giacommo}
 have been done to clarify the confinement 
mechanism on the basis of the idea of abelian projection of QCD
\cite{thooft2}.  
The abelian projection of QCD is to extract an abelian theory 
performing  a partial gauge-fixing.  
After an abelian projection,  $SU(3)$
QCD can be regarded as a $U(1)\times U(1)$ abelian gauge theory
with magnetic monopoles and electric charges. 'tHooft 
conjectured that the condensation 
of the abelian monopoles is the confinement mechanism in 
QCD\cite{thooft2}.
        
An effective $U(1)$ monopole action is derived from vacuum 
configurations in $SU(2)$ QCD 
after an abelian projection in a special gauge called 
maximally abelian (MA) gauge\cite{shiba1,shiba3,shiba6,shiba5}.
Entropy dominance over energy of the monopole loops, i.e.,
condensation of the monopole loops seems to occur always 
(for all values of the coupling constant $\beta$)
in the infinite-volume limit when extended 
monopoles\cite{ivanenko} are considered\cite{shiba1,shiba3,shiba6}. 
The abelian charge is confined due to the monopole condensation. 
The confinement of the abelian charge after abelian projection means 
color confinement as shown in \cite{thooft2}.
Monopole condensation is known to be the confinement mechanism 
also in lattice compact QED\cite{poly,bank,degrand,peskin,frolich,smit}. 

The string tension is a key quantity of confinement. 
It vanishes at the deconfinement transition temperature $T_c$\cite{gao}.
It was shown that 
the same string tension can be derived  from Wilson loops 
written in terms of abelian link fields alone after the abelian projection 
in the MA gauge\cite{suzu93,yotsu}. The string tension 
derived from the abelian Wilson loops also vanishes at $T_c$ \cite{ejiri}.
Moreover, the abelian Wilson loops can be expressed by a product of monopole 
and photon contributions\cite{shiba2,shiba4}. 
The monopoles alone are responsible 
for the string tension in $T=0$\cite{shiba2,shiba4} 
and $T\neq 0$\cite{ejiri} $SU(2)$ QCD. The same results are obtained  
also in $SU(3)$ QCD\cite{ilyar}. 

These results strongly support the 'tHooft conjecture\cite{thooft2} 
in $SU(2)$ QCD. 
However, the above results are 
restricted to the special MA gauge. What happens in other abelian 
projections is not known yet. For example, abelian Wilson loops in 
some unitary gauges are too small to derive reliably 
the string tension\cite{yotsu},
although it does not mean that the string tension can not be derived in the 
gauges. 

A Polyakov loop is another good order parameter, the vanishing of which  
means color-flux squeezing in the confinement phase. It is the aim of 
this note to show that 1)\ a Polyakov loop written in terms of abelian 
link fields alone is a good order parameter\cite{hio}, 2)\ it can be written 
by a product of contributions from Dirac strings of monopoles 
and from photons, 3)\ the former alone vanishes in the confinement phase 
and 4)\ these results are independent of the 
gauge choice of abelian 
projection. The first statement in the MA gauge
was already shown in \cite{hio}.   
 
\section{Dirac--string and photon contributions to Polyakov loops}
We adopt the usual $SU(2)$ Wilson action.
To study gauge dependence, we consider here three types of abelian projection,
i.e., the MA gauge and two unitary gauges. 
The MA  gauge is given \cite{kron} 
by performing a local gauge transformation $V(s)$ such that 
$$  R=\sum_{s,\mu}{\rm Tr}\Big(\sigma_3 
\widetilde{U}(s,\mu)
              \sigma_3 \widetilde{U}^{\dagger}(s,\mu)\Big)   $$
is maximized. Then a matrix 
\begin{equation}
X_1 (s)= \sum_{\mu}\Big(\widetilde{U}(s,\mu)
\sigma_3 \widetilde{U}^{\dagger}(s,\mu)
+\widetilde{U}^{\dagger}(s-\hat\mu,\mu)\sigma_3 
\widetilde{U}(s-\hat\mu,\mu)\Big) \label{x1}
\end{equation}
is diagonalized.
Here 
\begin{equation}
\widetilde{U}(s,\mu)=V(s)U(s,\mu)V^{-1}(s+\hat\mu). \label{vuv}
\end{equation}

Two unitary gauges considered here are defined by 
 performing a local gauge transformation $V(s)$ such that 
one of the following two matrices is diagonalized:
\begin{eqnarray}
X_2 (s) & = & \prod_{i=1}^{N_4}\widetilde{U}(s+(i-1)\hat4,4) 
\label{x2} \\
X_3 (s) & = & \sum_{\mu\neq\nu}
\widetilde{U}(s,\mu)\widetilde{U}(s+\hat\mu,\nu) 
\widetilde{U}^{\dagger}(s+\hat\nu,\mu)\widetilde{U}^{\dagger}(s,\nu),
\label{x3} 
\end{eqnarray}
where the sum in (\ref{x3}) is over all plaquette directions.
We call the former (the latter) Polyakov ($F12$) gauge.

After  the  gauge fixing is over, there still remains a $U(1)$ 
symmetry. We can extract an abelian link gauge 
variable  from the $SU(2)$ ones as follows;
\begin{equation}
   \widetilde{U}(s,\mu) =
        A(s,\mu)u(s,\mu), \label{au}       
\end{equation}
 where $u(s,\mu)$ is a diagonal abelian gauge field and  
$A(s,\mu)$ has off-diagonal components corresponding to
charged matters. Note that a $U(1)$ invariant 
quantity 
written in terms of the abelian link variables $u(s,\mu)$
 after an abelian projection is $SU(2)$ invariant\cite{shiba4,hio}.

Now let us show that an abelian Polyakov loop operator 
after the abelian projection 
\begin{eqnarray}
P = {\rm Re}[\exp\{i\sum_{i=1}^{N_4} J_4 (s+(i-1)\hat4)
\theta_4 (s+(i-1)\hat4)\}],
 \label{apol}
\end{eqnarray}
is given  by a product of monopole and photon contributions. 
Here  $J_4 (s)$ 
is an external current taking $+1$ along the straight line in the fourth 
direction 
 and $\theta_4 (s)$ is an angle variable defined from  $u(s,\mu)$ 
as follows:
\begin{eqnarray}
u(s,\mu)= \left( \begin{array}{cc}
e^{i\theta_{\mu}(s)} & 0 \\
0 & e^{-i\theta_{\mu}(s)} 
\end{array} \right).
\end{eqnarray}

Using the definition of a plaquette variable 
$f_{\mu\nu}(s)= \partial_{\mu}\theta_{\nu}(s) - 
\partial_{\nu}\theta_{\mu}(s)$ where $\partial_{\mu}$ is a 
forward difference, we get 
\begin{equation}
\theta_4 (s)= -\sum_{s'} D(s-s')[\partial'_{\nu}f_{\nu 4}(s')+
\partial_4 (\partial'_{\nu}\theta_{\nu}(s'))], \label{t4}
\end{equation} 
where $D(s-s')$ is the lattice Coulomb propagator 
and $\partial'_{\nu}$ is a backward difference. We have used 
$\partial_{\nu}\partial'_{\nu}D(s-s')=-\delta_{ss'}$.
Since $\partial'_4 J_4 (s) =0$, the second term in the right-hand side of 
(\ref{t4}) does not contribute to the abelian Polyakov loop (\ref{apol}).
Hence we get 
\begin{eqnarray}
P = {\rm Re}[\exp\{-i\sum_{i=1}^{N_4} J_4 (s+(i-1)\hat4)\sum_{s'}
D(s+(i-1)\hat4-s')\partial'_{\nu}f_{\nu 4}(s')\}].
 \label{apol2}
\end{eqnarray}

The gauge plaquette variable can be decomposed into two terms:
$$f_{\mu\nu}(s) = 
\bar{f}_{\mu\nu}(s)+ 2\pi n_{\mu\nu}(s),\ 
\ \ \ (-\pi < \bar{f}_{\mu\nu}(s) \le \pi) ,$$  
where 
$\partial'_{\mu}\bar{f}_{\mu\nu}(s)  $ includes only a photon 
field and $n_{\mu\nu}(s)$ 
is an integer-valued plaquette variable
denoting the number of Dirac strings through the plaquette 
coming out of monopoles\cite{degrand}. 
Hence we get 
\begin{eqnarray}
P & = & {\rm Re}[P_1 \cdot P_2],\\
P_1 & = &\exp\{-i\sum_{i=1}^{N_4} J_4 (s+(i-1)\hat4)\sum_{s'}
D(s+(i-1)\hat4-s')\partial'_{\nu}\bar{f}_{\nu 4}(s')\},\\
P_2 & = &\exp\{-2\pi i\sum_{i=1}^{N_4} J_4 (s+(i-1)\hat4)\sum_{s'}
D(s+(i-1)\hat4-s')\partial'_{\nu}n_{\nu 4}(s')\}.
\end{eqnarray}
We observe the photon ($P_{p}$) and the Dirac-string ($P_{m}$) 
contributions separately:
\begin{equation}
P_p = {\rm Re}[P_1] \hspace{1cm} {\rm and} \hspace{1cm} P_m = {\rm Re}[P_2].
\end{equation}

\section{The Villain form of QED} 
\begin{figure}
\epsfxsize=\textwidth
\begin{center}
\leavevmode
\epsfbox{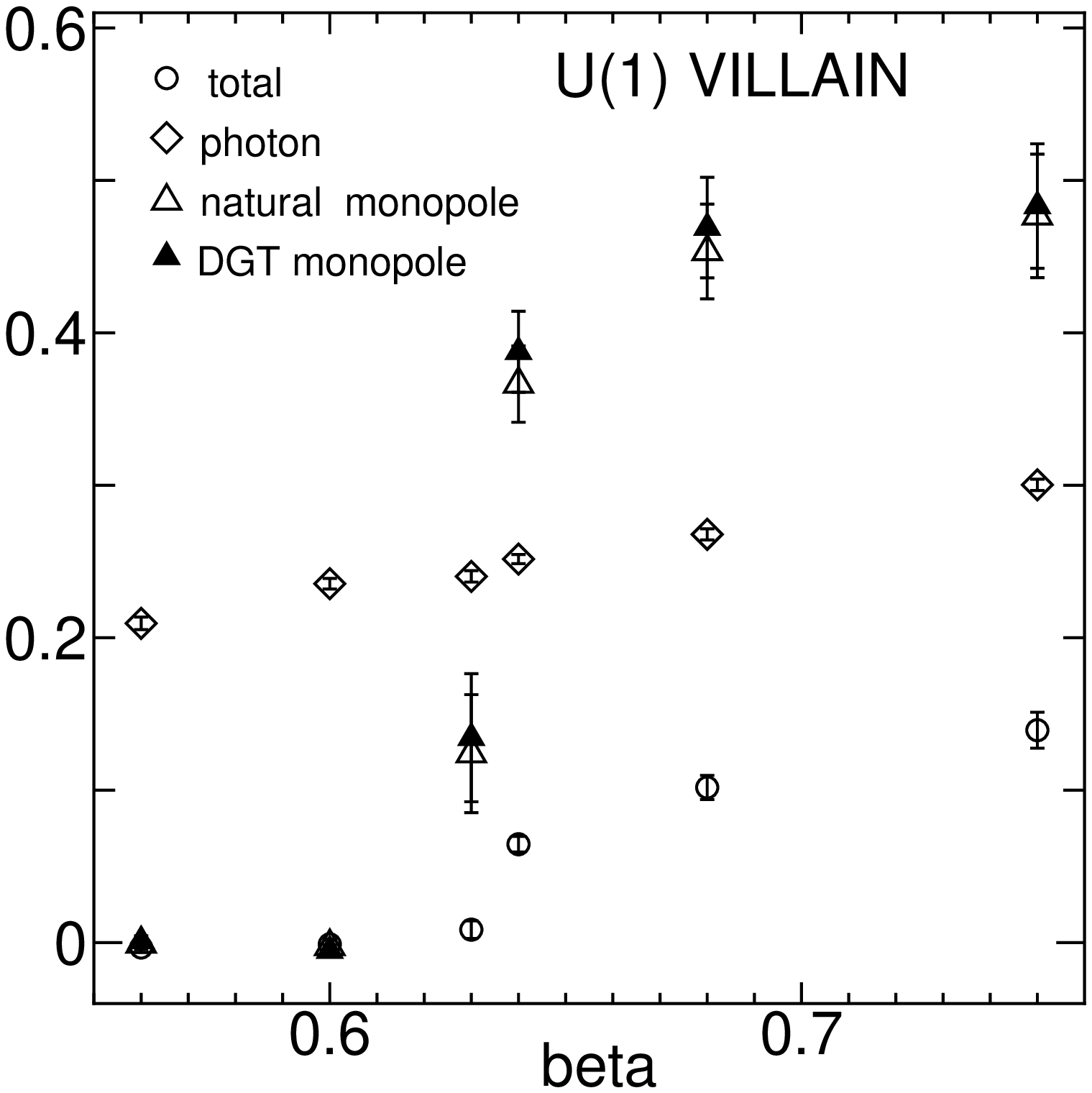}
\end{center}
\caption{
Monopole Dirac string and photon contributions to Polyakov loops in the 
Villain model of compact QED.
}
\label{villain}
\end{figure}

The above separation can be done also in the case of compact QED.
We first measure $P_p, P_m$ and $P$ adopting the Villain form\cite{villain}
 of the 
partition function on a $8^4$ lattice. Since there are natural monopoles 
and DeGrand-Toussaint monopoles in the Villain case of QED, we observe 
$P_m$ in terms of the two types of the monopoles.
Since the auto-correlation time is long for $\beta > \beta_c$, we have to 
perform Monte-Carlo simulations carefully. We follow the same method 
as done in \cite{shiba5}.
The results are shown in Fig.\ \ref{villain}. 
We find the following.
\begin{enumerate}
\item
The monopole Dirac-string data vanish in the confinement phase, 
whereas the photon data 
remain finite and change gradually for all $\beta$. 
The characteristic features of the Polyakov loops are then 
due to the behaviors of the Dirac-string contributions alone.
\item
Monopole Polyakov loops show more enhancement than the total ones 
for $\beta > \beta_c$. 
\item
Both types of monopoles give almost the same results.
\end{enumerate}

\section{The MA gauge  in $SU(2)$ and $SU(3)$ QCD}
The Monte-Carlo simulations were done in $SU(2)$ 
on $16^3\times 4$ lattice 
from $\beta =
2.1$ to $\beta =2.5$ in the MA gauge and 
in  the unitary gauges. In $SU(3)$ QCD, we adopted $10^3 \times 2$ 
lattice from $\beta = 5.07$ to $\beta= 5.12$.
 All measurements were done every 50 sweeps (40 sweeps in the $SU(3)$ case) 
 after
a thermalization of 2000 sweeps. We took 50 
configurations totally for
measurements. The gauge-fixing criterion in the MA gauge 
is the same  as done in Ref.\ \cite{ohno}. 

\begin{figure}
\epsfxsize=\textwidth
\begin{center}
\leavevmode
\epsfbox{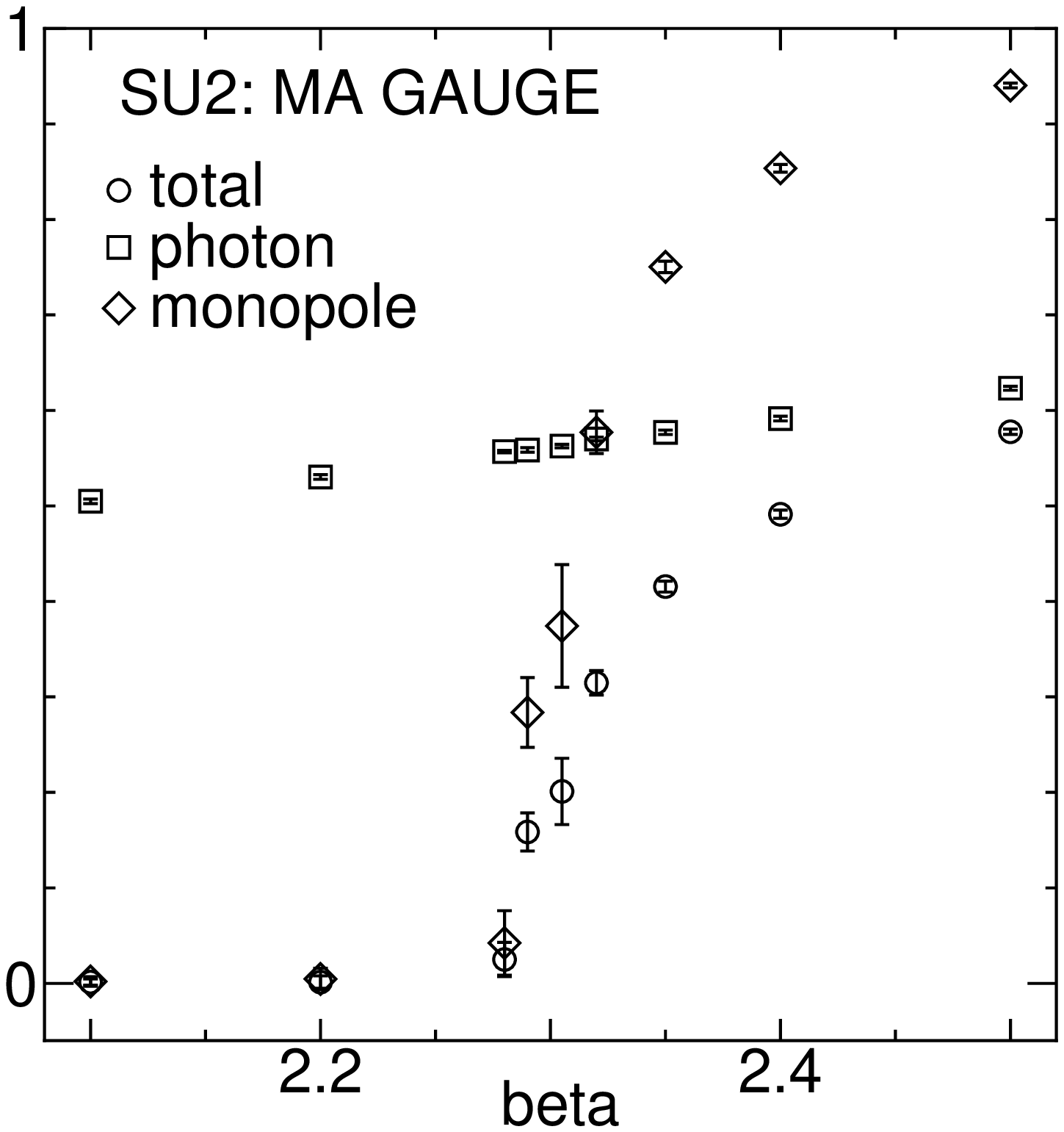}
\end{center}
\caption{
Monopole Dirac string and photon contributions to Polyakov loops in the 
MA gauge in $SU(2)$ QCD.
}
\label{ma}
\end{figure}

\begin{figure}
\epsfxsize=\textwidth
\begin{center}
\leavevmode
\epsfbox{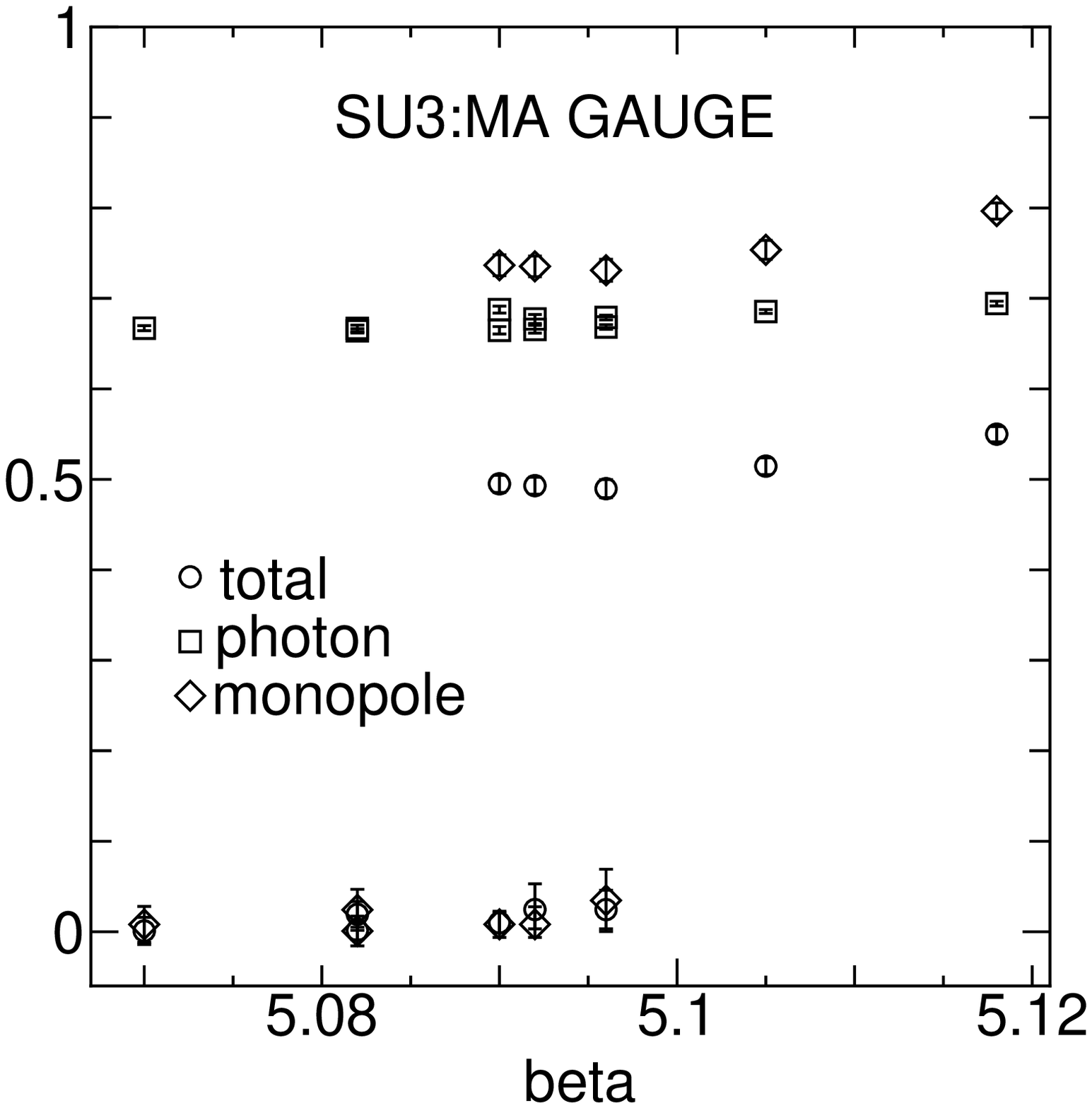}
\end{center}
\caption{
Monopole Dirac string and photon contributions to Polyakov loops in the 
MA gauge in $SU(3)$ QCD.
}
\label{su3}
\end{figure}

The results in the MA gauge are shown in the following:
\begin{enumerate}
\item
We plot the $SU(2)$ data in the MA gauge in Fig.\ \ref{ma}.
The abelian Polyakov loops remain zero in the confinement phase, whereas 
they begin to rise from the critical temperature $\beta_c = 2.298$\cite{satz}
This was observed already in \cite{hio}. It is interesting that 
the Dirac-string contribution shows similar behaviors more drastically. 
It is zero for $\beta < \beta_c$, whereas it begins to rise rapidly and it 
reaches $\sim 1.0$ for large $\beta$. On the other hand, the photon 
part has a finite contribution for both phases and it changes only slightly.
Characteristic behaviors of the abelian Polyakov loops as an 
order parameter of deconfinement transition are then explained 
by the Dirac-string part of monopoles alone. This is consistent with the 
results in \cite{shiba2,shiba4,ejiri} 
stating monopoles alone are responsible for 
the value of the string tension. 
\item
The same results are obtained also in 
pure $SU(3)$ QCD in the MA gauge as shown in Fig.\ \ref{su3}. 
The monopole Dirac string alone is seen to be responsible for the 
flux squeezing. There is 
a clear hysteresis behavior showing the first order transition.
\end{enumerate}

\section{The unitary gauges}
\begin{figure}
\epsfxsize=\textwidth
\begin{center}
\leavevmode
\epsfbox{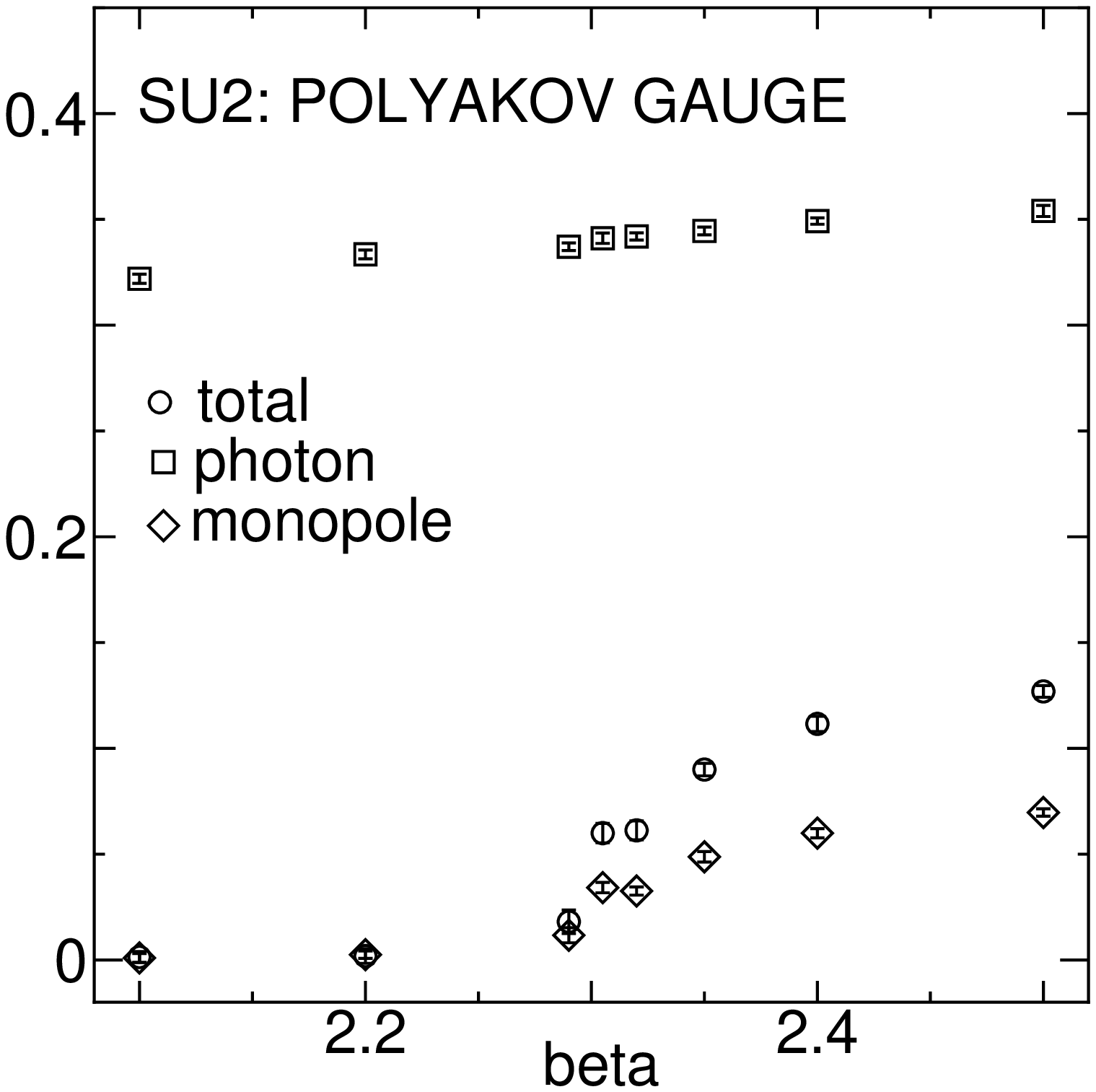}
\end{center}
\caption{
Monopole Dirac string and photon contributions to Polyakov loops in the 
Polyakov gauge.
}
\label{pol}
\end{figure}

\begin{figure}
\epsfxsize=\textwidth
\begin{center}
\leavevmode
\epsfbox{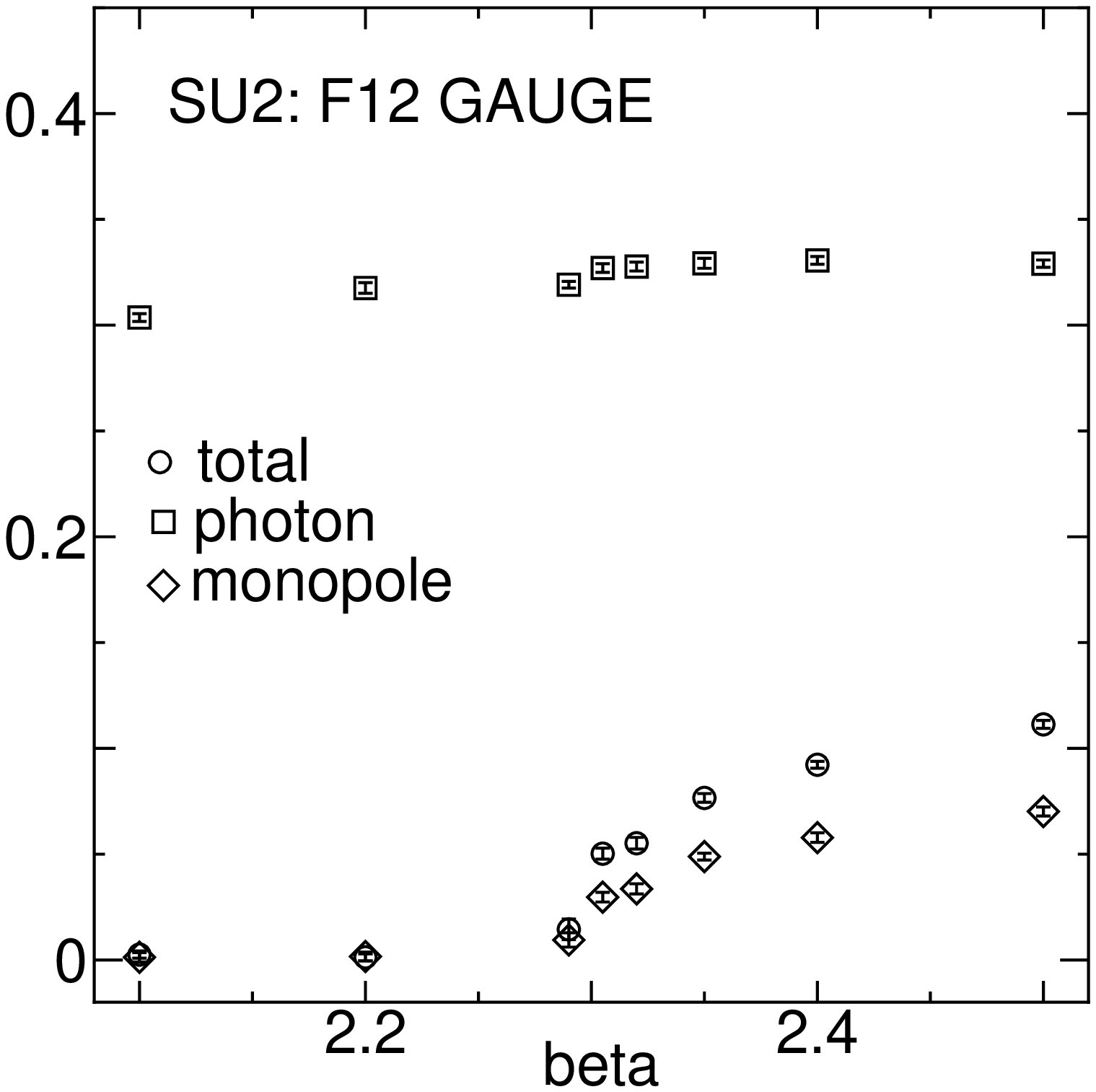}
\end{center}
\caption{
Monopole Dirac string and photon contributions to Polyakov loops in the 
$F12$ gauge.
}
\label{f12}
\end{figure}

\begin{figure}
\epsfxsize=\textwidth
\begin{center}
\leavevmode
\epsfbox{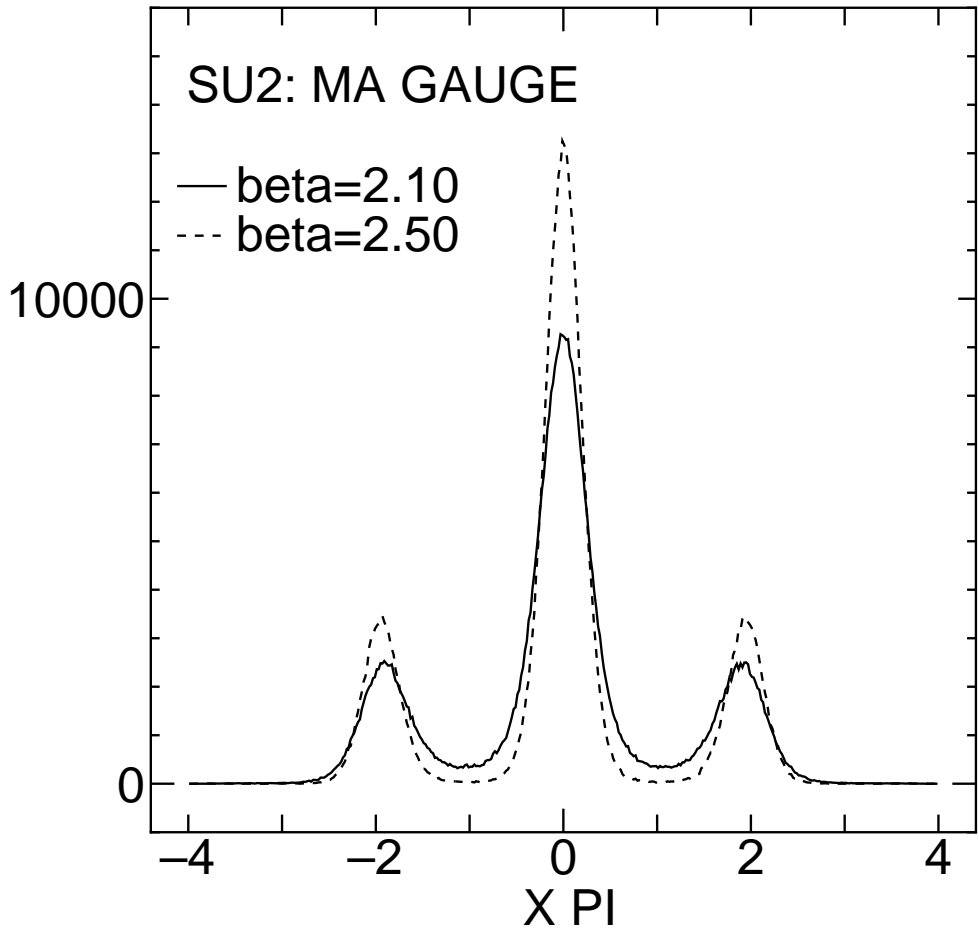}
\end{center}
\caption{
Histogram of $f_{\mu\nu}$ in the MA gauge for some $\beta$.
}
\label{fmunu}
\end{figure}

\begin{figure}
\epsfxsize=\textwidth
\begin{center}
\leavevmode
\epsfbox{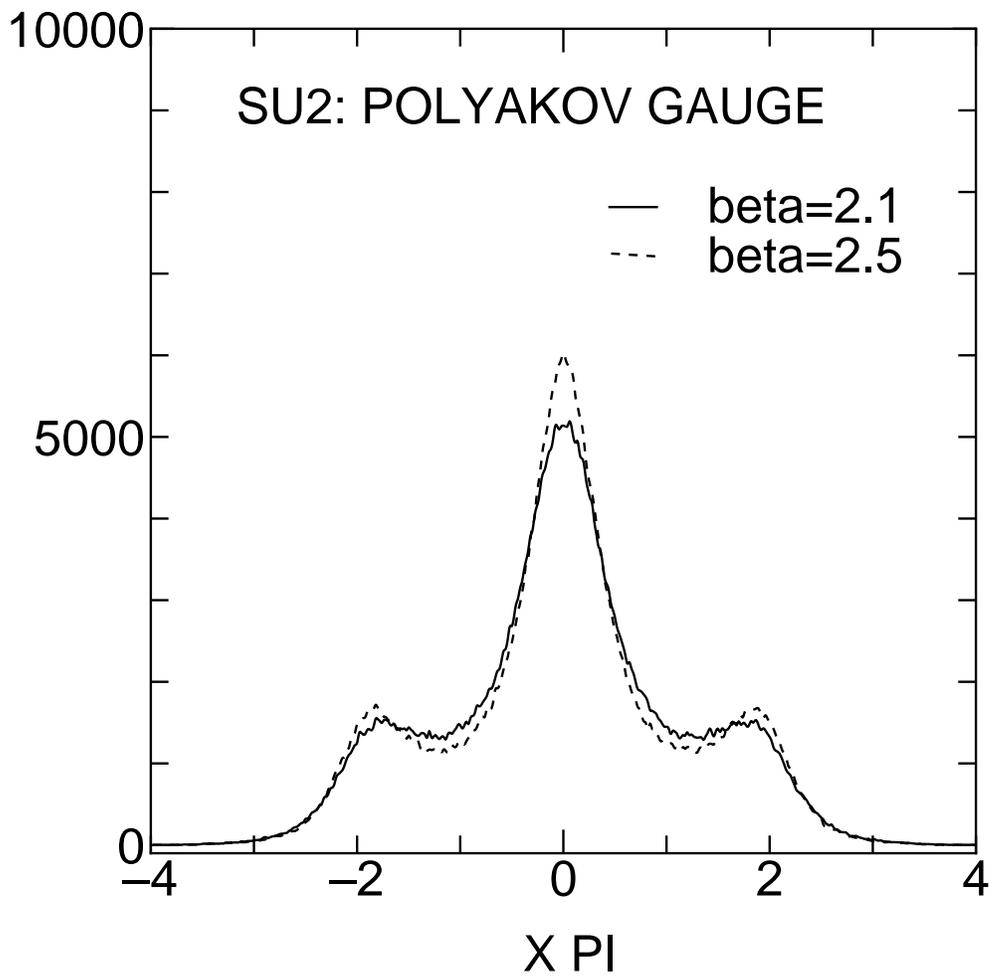}
\end{center}
\caption{
Histogram of $f_{\mu\nu}$ in the Polyakov gauge for some $\beta$.
}
\label{fmunup}
\end{figure}

We next study the case in the unitary gauges. 
\begin{enumerate}
\item
The data in the unitary gauges are plotted in Fig.\ \ref{pol} 
and Fig.\ \ref{f12}. 
It is very interesting to see that qualitative features are similar to those 
in the MA gauge. Namely, the abelian and the Dirac-string Polyakov loops are 
zero in the confinement phase, which suggests occurrence of flux squeezing 
in the unitary gauges, too. They show finite contribution above the critical 
temperature $\beta_c$. Photon contributions are finite and change 
 gradually in both phases. The Dirac-string part of monopoles 
is responsible for the 
essential features of an order parameter. 
{\it These are the first phenomena suggesting gauge independence of the 
'tHooft conjecture.} 

\item
Comparing the figures in the unitary gauges, one can see both are almost 
equal, although the gauge fixing conditions are quite different. 
The finite values in the deconfinement phase in Fig.\ \ref{pol}
 and Fig.\ \ref{f12} are much smaller than those 
in Fig.\ \ref{ma} in the MA gauge. 
It is similar to the behaviors of Wilson loops in these gauges studied 
in \cite{yotsu}.  Namely, abelian Wilson loops in the MA gauge enhance 
drastically and then we could determine the string tension. On the other hand, 
the Wilson loops in the unitary gauges are too small to fix the string tension
reliably.  

\item
There is a problem in extracting monopoles 
in the unitary gauges in comparison 
with the case in the MA gauge, as seen from the histogram of 
$f_{\mu\nu}$ in some configurations. The data in both gauges are shown in 
Fig.\ \ref{fmunu} and Fig.\ \ref{fmunup}. 
The data in the MA gauge was first studied
 in \cite{shiba7}. 
In the MA gauge, quantum fluctuation is small and it is reliable to 
separate  Dirac strings out of  $f_{\mu\nu}$. However, in the 
unitary gauges, the separation 
between Dirac strings and quantum fluctuation is more ambiguous as seen from 
the bump around $\pm \pi$ in the histogram. To be noted, the bump becomes 
smaller as $\beta$ becomes larger. This is because quantum fluctuation becomes 
smaller for larger $\beta$. Hence, we may expect that 
the correct string tension can be derived 
even in unitary gauges when we consider the $T = 0$ case for large $\beta$
on larger lattices.
\end{enumerate}

\section{Conclusion and remarks}
In conclusion,  our analyses done here strongly suggest 
that abelian monopoles are responsible for 
confinement in $SU(2)$ QCD 
and condensation of the monopoles is the confinement mechanism. 
We have found first the data suggesting gauge independence of the 'tHooft 
conjecture.  

Our data show the following picture of color confinement due to monopole 
condensation. Choose any $U(1)$ out of $SU(2)$ through an abelian projection.
Then quarks and gluons behave like charged matters with respect to the $U(1)$ 
symmetry after the abelian projection. There are always monopoles 
with magnetic charges with respect to the magnetic $U(1)$ symmetry dual to the 
$U(1)$ chosen. The electric $U(1)$ charge is confined
 due to the dual Meissner effect 
caused by the condensation of the corresponding monopoles. 
The electric charge confinement after  
abelian projection is equivalent to color confinement as proved in 
\cite{thooft2}. Gauge independence of the confinement mechanism appears in 
this way.

\begin{figure}
\epsfxsize=\textwidth
\begin{center}
\leavevmode
\epsfbox{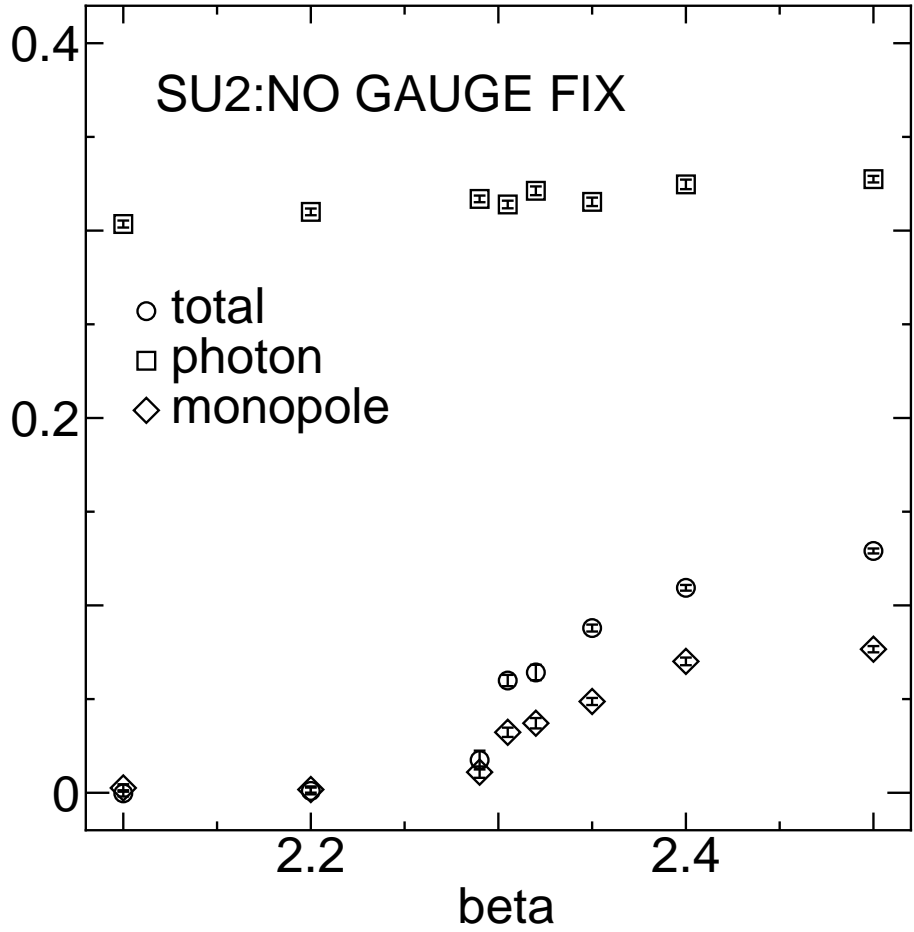}
\end{center}
\caption{
Monopole Dirac string and photon contributions to abelian Polyakov loops 
without gauge-fixing. 
}
\label{ngpol}
\end{figure}

\begin{figure}
\epsfxsize=\textwidth
\begin{center}
\leavevmode
\epsfbox{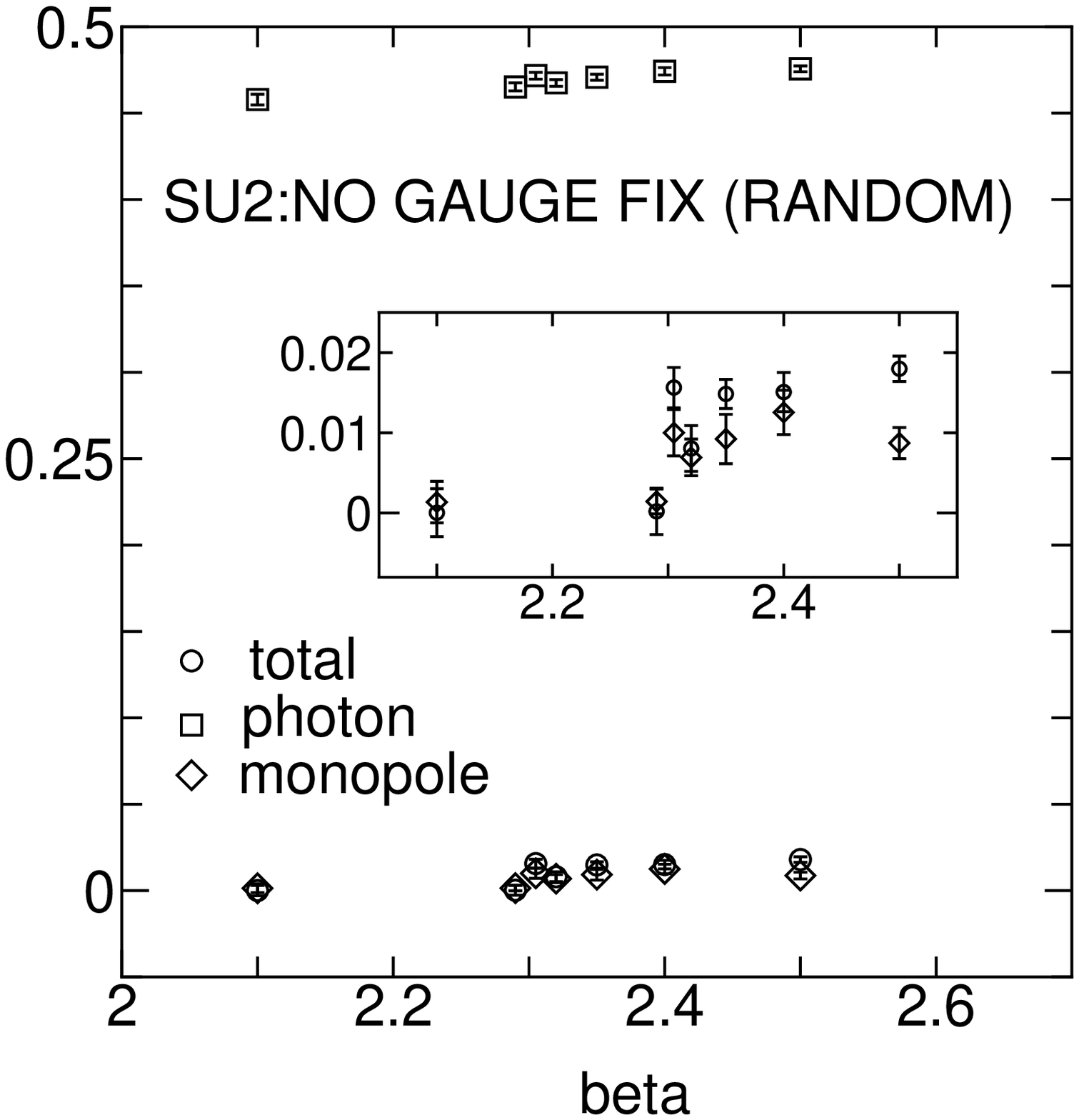}
\end{center}
\caption{
Monopole Dirac string and photon contributions to abelian Polyakov loops 
without gauge-fixing. Abelian links are defined randomly at each site.
}
\label{ngpolr}
\end{figure}

In \cite{yotsu}, abelian Wilson loops have been measured in various gauges.
The abelian Wilson loops in the unitary gauges do not show abelian 
dominance and take a similar value to that without gauge-fixing. 
Suggested from the study, we have tried to measure abelian Polyakov loops 
without gauge-fixing, although such quantities are gauge variant. 
Abelian link fields are defined by choosing any one of isospin directions.
Surprisingly enough, we have obtained a similar behavior 
as shown in Fig.\ \ref{ngpol}. The abelian Polyakov loop is zero 
in the confinement phase and shows rising at the critical $\beta_c$.
The responsibility of the Dirac string is also seen. 
The abelian quantity without gauge-fixing is variant under local 
$SU(2)$ transformation. Hence to see whether the above behavior includes some 
physical meaning or not, we have also measured abelian Polyakov loops 
composed of abelian links defined randomly at each site. Namely  abelian 
links in the time direction at each site are defined to take a random isospin 
direction. The data are shown in Fig.\ \ref{ngpolr}. 
Qualitatively similar behaviors are obtained, although the finite 
values of the total and the Dirac-string contributions 
in the deconfinement region  are smaller. 
These data may support the above picture of gauge independence of the 
'tHooft idea.  To prove gauge independence  definitely is to be 
studied in future.

This work is financially supported by JSPS Grant-in Aid for 
Scientific  Research (B)(No.06452028).

\input epsf

\begin{references}
\bibitem{suzu93} T. Suzuki, Nucl. Phys. B(Proc. Suppl.) 
{\bf 30}, (1993) 176 and see also references therein.  
\bibitem{shiba1} H.Shiba and T.Suzuki, 
Kanazawa University Report KANAZAWA 93-09, 1993 (unpublished). 
\bibitem{shiba2} H.Shiba and T.Suzuki, 
Kanazawa University Report KANAZAWA 93-10, 1993 (unpublished). 
\bibitem{shiba3} H.Shiba and T.Suzuki, 
Nucl. Phys. B(Proc. Suppl.) {\bf 34}, (1994) 182 .
\bibitem{cosmai} P.Cea and L.Cosmai,
Nucl. Phys. B(Proc. Suppl.) {\bf 34}, (1994) 219.
\bibitem{yee} K.Yee,
Nucl. Phys. B(Proc. Suppl.) {\bf 34}, (1994) 189.
\bibitem{polika} M.N.Chernodub, M.I.Polikarpov and M.A.Zubkov,
Nucl. Phys. B(Proc. Suppl.) {\bf 34}, (1994) 256.
\bibitem{wensley} R.J.Wensley,
Nucl. Phys. B(Proc. Suppl.) {\bf 34}, (1994) 204.
\bibitem{shiba4} H.Shiba and T.Suzuki, 
Kanazawa University Report KANAZAWA 94-07, 1994, 
to be published in Physics Letters B. 
\bibitem{shiba6} H.Shiba and T.Suzuki, 
Kanazawa University Report KANAZAWA 94-12, 1994. 
\bibitem{ejiri} S.Ejiri, S.Kitahara, Y.Matsubara and T.Suzuki, 
Kanazawa University Report KANAZAWA 94-14, 1994. 
\bibitem{stack} J.D.Stack, S.D.Neiman and R.J.Wensley,
University of Illinois preprint, ILL-(TH)-94-\#14, 1994.
\bibitem{giacommo} L. Del Debbio $et\  al.$, University of Pisa report 
IFUP-TH 16/94, 1994.
\bibitem{thooft2} G. 'tHooft, Nucl. Phys. {\bf B190}, (1981) 455.
\bibitem{shiba5} H.Shiba and T.Suzuki, 
Kanazawa University Report KANAZAWA 94-11, 1994. 
\bibitem{ivanenko} T.L. Ivanenko $et\  al.$, Phys. Lett. {\bf B252},
 (1990) 631.
\bibitem{poly} A.M. Polyakov, Phys. Lett. {\bf B59}, (1975) 82.
\bibitem{bank} T.Banks $et\  al.$, Nucl. Phys. {\bf B129}, (1977) 493.
\bibitem{degrand} T.A. DeGrand and D. Toussaint, 
Phys. Rev. {\bf D22}, (1980) 2473.
\bibitem{peskin} M.E. Peshkin, Ann. Phys. {\bf 113}, (1978) 122.
\bibitem{frolich} J. Fr\"{o}lich and P.A. Marchetti, Euro. Phys. Lett.
{\bf 2}, (1986) 933.
\bibitem{smit} J. Smit and A.J. van der Sijs, Nucl. Phys. 
{\bf B355}, (1991) 603; Nucl. Phys. B(Proc. Suppl.) 
{\bf 20}, (1991) 221.
\bibitem{gao}M. Gao, Nucl. Phys. B(Proc. Suppl.){\bf 9}, (1989) 368.
\bibitem{yotsu} T. Suzuki and I. Yotsuyanagi, 
Phys. Rev. {\bf D42}, (1990) 4257.
\bibitem{ilyar} S. Ilyar $et\  al.$, in preparation.
\bibitem{hio} S. Hioki $et\  al.$, Phys. Lett. {\bf B272}, (1991) 326.
\bibitem{kron} A.S. Kronfeld $et\  al.$, Phys. Lett. {\bf B198}, 
(1987) 516;
A.S. Kronfeld $et\  al.$, Nucl. Phys. {\bf B293}, (1987) 461.
\bibitem{villain}J. Villain, J. Phys. (Paris) {\bf 36}, (1975) 581.
\bibitem{ohno} S.Hioki $et\  al.$,Phys.Lett. {\bf B271}, (1991) 201.  
\bibitem{satz} J.Engels, J.Jersak, K.Kanaya, E.Laermann, C.B.Lang,
T.Neuhaus and H.Satz,
Nucl. Phys. {\bf B280},(1987) 577.
\bibitem{shiba7} H. Shiba, in preparation.
\end{references}
\end{document}